\documentclass{article}
\usepackage{ijcai21}

\usepackage{times}
\usepackage{soul}
\usepackage{url}
\usepackage[hidelinks]{hyperref}
\usepackage[utf8]{inputenc}
\usepackage[small]{caption}
\usepackage{graphicx}
\usepackage{amsmath}
\usepackage{amsthm}
\usepackage{amssymb}
\usepackage{booktabs}
\usepackage{algorithm}
\usepackage{algorithmic}
\usepackage{xcolor}

\title{AutoFT: Automatic Fine-Tune for Parameters Transfer Learning in Click-Through Rate Prediction}
\author{
Xiangli Yang$^1$ \footnote{This work was done when Xiangli Yang was an intern at Huawei Noah’s Ark Lab.},
Qing Liu$^2$,
Rong Su$^2$,
Ruiming Tang$^2$,
Zhirong Liu$^2$,
Xiuqiang He$^2$
\affiliations
$^1$Chongqing Jiaotong University\\
$^2$Huawei Noah’s Ark Lab
\emails
xiangli.yang@cqjtu.edu.cn,
\{liuqing48,surong3,tangruiming,liuzhirong,hexiuqiang1\}@huawei.com
}

\usepackage{multirow}
\usepackage{caption}

\begin{document}

\maketitle

\begin{abstract}
Recommender systems are often asked to serve multiple recommendation scenarios or domains. Fine-tuning a pre-trained CTR model from source domains and adapting it to a target domain allows knowledge transferring. However, optimizing all the parameters of the pre-trained network may result in over-fitting if the target dataset is small and the number of parameters is large. This leads us to think of directly reusing parameters in the pre-trained model which represent more general features learned from multiple domains. However, the design of freezing or fine-tuning layers of parameters requires much manual effort since the decision highly depends on the pre-trained model and target instances. In this work, we propose an end-to-end transfer learning framework, called Automatic Fine-Tuning (AutoFT), for CTR prediction. AutoFT consists of a field-wise transfer policy and a layer-wise transfer policy. The field-wise transfer policy decides how the pre-trained embedding representations are frozen or fine-tuned based on the given instance from the target domain. The layer-wise transfer policy decides how the high-order feature representations are transferred layer by layer. Extensive experiments on two public benchmark datasets and one private industrial dataset demonstrate that AutoFT can significantly improve the performance of CTR prediction compared with stateof-the-art transferring approaches.
\end{abstract}

\section{Introduction}\label{sec:intro}

Modern recommender systems are widely deployed to make precise personalized recommendations. In recommender systems, Click-Through Rate (CTR) prediction is a crucial task, which estimates the probability that a user will click on a recommended item under a specific context, so that the recommendation decisions can be made based on the predicted CTR values~\cite{DBLP:conf/recsys/Cheng0HSCAACCIA16/wideanddeep,DBLP:conf/ijcai/GuoTYLH17/deepfm,qu2016product/pnn,qu2018product/ipnn,DBLP:conf/kdd/LianZZCXS18/xdeepfm,DBLP:conf/kdd/WangFFW17}. Recommender systems are often asked to serve multiple recommendation scenarios. However, designing a unique CTR prediction model for each scenario (i.e., domain or task) is hard to achieve due to the limit of computing resource and human resource.
%training a deep model requires a lot of manual work to find a set of proper hyper parameters of a model.
Making it even worse, training a model on a domain with little data often leads to over-fitting. A common practical solution is to integrate data from different domains but with similar distributions and train a ``general'' model.
Then this ``general'' model is deployed to serve different scenarios.
The drawback of this solution is obvious, i.e., this ``general'' model is not the optimal for every and each target domain.

\begin{figure}
  \centering
    \includegraphics[width=0.48\textwidth]{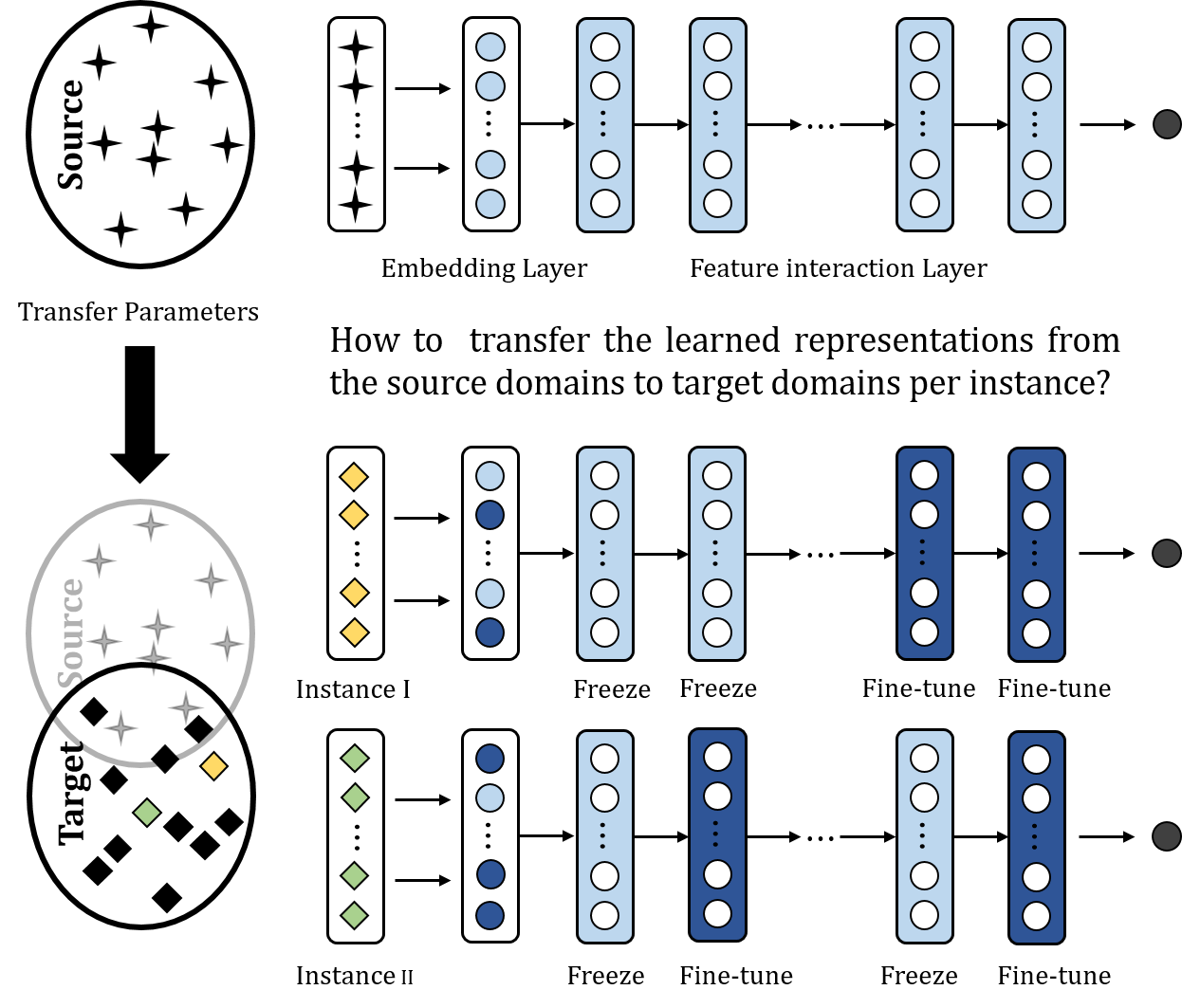} 
    \caption{Introduction of parameters transfer learning of CTR model. This is an example showing that different instances in the target domain may need different amount of source information to fine-tune the model parameters.
    } 
    \label{fig:intro1} 
\end{figure}

Transfer learning is a promising way of helping train a more precise model for each target domain.
In this work, we deal with the task of transferring the ``general'' CTR model trained from the source domains to a target domain\footnote{We focus on the case of only one target domain. When there exist multiple target domains, we transfer the ``general'' CTR model to each target domain individually.}. 
In particular, we focus on deep CTR models as the ``general'' model, pre-trained in a supervised manner on the source domains.
% we attempt to use deep neural network models, pre-trained
% in a supervised manner on source domains with rich user-item interactions, for a variety of tasks
% on target domains to improve the prediction performance of the target domain. 
To do so, we need to tackle the following challenges.

(C1) Existing recommendation literature is unclear on whether deep CTR models can be transferred from source domains to target domains where different domains may not share users and items. Moreover, there are various deep CTR models in the literature. Therefore, a transfer learning framework that is compatible with any deep CTR model is needed. 
% Develop an architecture that can transfer pre-trained deep CTR prediction models to downstream domains. 
(C2) Deep CTR models follow an \emph{embedding} \& \emph{feature interaction} paradigm. It is hard to determine whether and how each of these two modules are transferred to a target domain, so that the model performance on the target domain is superior. 
% it is most  which components of a deep CTR model, e.g., the embedding layer (for feature representation), the deep layers or the cross layers (for feature interaction), of a CTR prediction model should be transferred and how they are transferred so that it is profitable for the target domain.
(C3) Different training instances and features may need different amount of information from source domains as shown in Figure \ref{fig:intro1}. For example, items that are less frequent in the target domain may need more information from the source domains than the more popular items in the target domain. Such fine granularity control of knowledge transfer is hardly possible to manage by experts.

Transfer learning and domain adaptation techniques are well studied in computer vision (CV) area \cite{yosinski2014transferable,rusu2016progressiveNN,xuhong2018explicit/L2SP,guo2019spottune} and natural language processing (NLP) area \cite{DBLP:conf/naacl/DevlinCLT19,DBLP:conf/naacl/RuderPSW19}, but not much in CTR prediction area. 
The typical way of performing transfer learning with deep neural networks is to fine-tune all the parameters of a pre-trained model on the source domain using data from the target domain.
However, it is unclear whether fine-tuning all the parameters for all the instances in the target domain is the optimal solution.
These works \cite{yosinski2014transferable,rusu2016progressiveNN,xuhong2018explicit/L2SP,guo2019spottune} proposed in CV area suggest to import the pre-trained model information wisely. For example, ProgressiveNN (Progrssive Neural Network) \cite{rusu2016progressiveNN} proposes to combine the parameters of the pre-trained network and the fine-tuned network via MLP.
$L^2$-SP \cite{xuhong2018explicit/L2SP} investigates several regularization schemes and recommends a simple $L^2$ penalty to encourage the similarity of the pre-trained network and the fine-tuned network. 
SpotTune \cite{guo2019spottune} treats instances differently during fine-tuning by training a policy network to decide which layer of a model should be frozen or fine-tuned for individual instances. 
% These methods solve the first challenge mentioned above by giving us a general idea of how to re-use the parameters in a pre-trained network to improve training of the target domain.
However, these works cannot resolve C2 and C3 if we apply them directly in CTR prediction area, since they either do not distinguish different components of a deep CTR model (such as \cite{guo2019spottune}) or do not consider different network layers and individual instances (like \cite{xuhong2018explicit/L2SP,rusu2016progressiveNN}) during the transfer learning. 
%Thus, they can not solve challenge (2) and (3).
More importantly, these models are designed for CV where the input are continuous values that represent quantifiable values of images, which is quite different from CTR prediction area.

% It is unclear how these works perform for CTR prediction where the feature representation and feature interaction play important roles in the prediction.

In this paper, we explore how transferable are parameters in a pre-trained CTR prediction model, which usually consists of an embedding layer and a feature interaction layer, from multiple source domains to a target domain. 
We mainly focus on transferring the parameters of a pre-trained model based on source domains with proper fine-tuning.  
Other transfer learning approaches such as instance based approaches~\cite{lin2013double/instancebaesd1,zhang2014instancebased2,wang2019minimax/instancebased3,DBLP:conf/kdd/Zhong0YBH20}, feature based approaches~\cite{pan2010featurebased1,pan2008transfer/featurebased2} and domain adaptation approaches~\cite{ganin2015unsupervised/da1,long2016unsupervised/da2,aggarwal2019domain/da3,tran2019domainadapt/da4,DBLP:conf/www/LiuZZLSX0X20} are beyond the scope of this work.
%These approaches will be considered as future work.

We propose an adaptive fine-tuning approach for transferring learning on deep CTR models,
% develop our method based on a widely used ``deep and cross", i.e., DCN , CTR prediction model by proposing an adaptive fine-tuning approach, 
called AutoFT, which automatically finds a route between the pre-trained network and the siamese fine-tuning network per instance in the target domain. Exemplary routes are depicted in red color in Figure \ref{fig:framework} .
In particular, to cope C1, AutoFT is a general framework to import parameters from the pre-trained model, which is compatible to various deep CTR models in the literature. Without lose of generality, we use DCN~\cite{DBLP:conf/kdd/WangFFW17} model to elaborate the details of AutoFT.
% refer to the stat-of-the-art approaches \cite{rusu2016progressiveNN,xuhong2018explicit/L2SP,guo2019spottune} and explored three three classic ways, i.e., linearly add up, regularization, and initialization, of importing parameters from the pre-trained model to the new model.
For C2, different transfer strategies are designed in AutoFT for different components of a deep CTR model, since different components play unique roles in CTR prediction. Specifically, we propose a \emph{field-wise} transfer policy for the embedding layer and a \emph{layer-wise} transfer policy for the feature interaction layers. 
%Our method can be adapted to many deep learning based CTR models.
To handle C3, we design a set of instanced-based policy networks to decide which parameters should be preserved as in the pre-trained model (frozen) or fine-tuned, so that feasible parameter adaptation for different instances and features is achieved.
%to feasibly adapt to different level of ``general" information.

In summary, we make the following contributions in this paper. 
\begin{itemize}
    \item Firstly, we propose a framework AutoFT for transferring parameters of a pre-trained CTR prediction model from source domains to a target domain. AutoFT automatically finds a route between the pre-trained network and the siamese fine-tuning network per instance in the target domain and is compatible with any deep CTR models. 
    % To the best of our knowledge, this is the first framework that explores the transferablity of the parameters of CTR prediction models to serve different recommendation scenarios.
    
    \item We propose a field-wise transfer policy for the embedding layer, and a layer-wise transfer policy for the feature interaction layers. Feasible parameter adaption for different features and instances is achieved by designing a set of learnable policy networks. 
    
    \item We compare the performance of our propose approach with many state-of-the-art methods which is also designed for re-using the parameters of pre-trained models, on two public benchmarks and one private dataset. The experimental results demonstrate the superiority of our proposed AutoFT in transfer learning. 
    
    \item We also study which components or layers of a CTR model need more fine-tuning. The result shows that there is similar phenomenon to computer vision (CV) that the lower layers of the network represent more general features while higher layers need more fine-tuning to fit for a specific target domain.

\end{itemize}

\section{Related Work}

\subsection{Deep Models in CTR prediction}
Recently, deep learning based models achieve state-of-the-art
performance for CTR prediction~\cite{DBLP:conf/recsys/Cheng0HSCAACCIA16/wideanddeep,DBLP:conf/ijcai/GuoTYLH17/deepfm,qu2016product/pnn,DBLP:conf/kdd/WangFFW17,DBLP:conf/kdd/LianZZCXS18/xdeepfm}. 
Deep CTR models follow an \emph{embedding} \& \emph{feature interaction} paradigm. Many works focus on designing new network architectures to improve the feature interaction component, aiming to better capture the nonlinear relationship among features.
%Deep learning based CTR prediction models expect Multi-layer Perceptions (MLP) to capture higher order feature interactions to improve prediction accuracy.

Wide \& Deep~\cite{DBLP:conf/recsys/Cheng0HSCAACCIA16/wideanddeep} proposed by Cheng et al. jointly trains a wide linear model for manually designed second-order feature interactions and a deep neural network for higher-order feature interactions. DeepFM~\cite{DBLP:conf/ijcai/GuoTYLH17/deepfm} and PNN~\cite{qu2016product/pnn} use an FM layer to learn second-order feature interactions and MLP to learn higher-order feature interactions.
% substitutes the wide component in Wide \& Deep with an FM layer which aims to model second-order feature interactions automatically. PNN~\cite{qu2016product/pnn} uses an FM layer to model second-order feature interactions, while followed by MLP to model the further interactions.
PIN~\cite{qu2018product/ipnn} introduces a network-in-network architecture to model pairwise feature interactions with sub-networks rather than using simple inner product operations as in PNN and DeepFM. DIN~\cite{zhou2018deep/DIN} and AFM~\cite{xiao2017attentional/AFM} utilize attention network to differentiate contribution of each second-order feature interactions.
DCN~\cite{DBLP:conf/kdd/WangFFW17} and xDeepFM \cite{DBLP:conf/kdd/LianZZCXS18/xdeepfm} apply a cross operation and a compressed interaction respectively, to explicitly learn $p^{\text{th}}$ order feature interactions from $(p-1)^{\text{th}}$ order ones.
% has designed a novel cross network to learn bounded-degree feature
% interactions and xDeepFM \cite{DBLP:conf/kdd/LianZZCXS18/xdeepfm} proposes to learn feature interactions in an vector-wise fashion.

These works mainly focus on designing network architectures for the feature interaction component of deep CTR models. They achieve good performance in a single domain.
In real-world applications, a well trained model in one or multiple source domains may be useful for a target domain, because sharing such information may be beneficial especially when the target domain is of limited data. In this paper, we study this problem and propose a transfer  learning framework for transferring parameters of a pre-trained CTR model from source domains to a target domain which is compatible with the above mentioned deep CTR models.

% many cases, we wish the model trained for one domain could be applied to another domain since it is expensive, with respect to time and resources, to train a large scale deep learning based CTR prediction model from the beginning.
% However, it is not trivial to transfer a model trained on one application domain to another since it is unclear whether the transferring is positive or negative.
% In this paper, we hope to find a flexible transferring strategy such that the information from the source domain could be positively reused for prediction tasks in target domain.

\subsection{Transfer Learning}

Transfer learning and domain adaptation techniques are widely used in computer vision while they are relatively rarely explored in the CTR prediction area.
For example, 
Yosinski et al. ~\cite{yosinski2014transferable} experimentally quantify the generality versus specificity of neurons in each layer of a deep convolutional neural network. They report 
that initializing a network with transferred features from almost any number of layers can boost the generalization that lingers even after fine-tuning to the target domain.
This result has yet been verified in CTR prediction area where the input feature is not usually continuous, and the model structure is quite different from computer vision.

While fine-tuning in~\cite{yosinski2014transferable} incorporates the prior knowledge only by initializing with a pre-trained model, ProgressiveNN~\cite{rusu2016progressiveNN} retains the parameters of the pre-trained model and learns another set of parameters from the pre-trained model via fine-tuning. Then, the pre-trained parameters and fine-tuned parameters are combined via MLP, which is proven to be effective for tansfer learning.
% lateral connections from these parameters throughout training by integrating with linear transformation at each layer of the feature hierarchy, which is proven to be effective for the new tasks.
$L^2$-SP~\cite{xuhong2018explicit/L2SP} assumes that the pre-trained model from source domains are useful for the target model.
% designed under the assumption that the pre-trained model extracts generic features, which
% are at least partially relevant for solving the target task.
This work explores several regularization schemes that
explicitly promote the similarity between the fine-tuned model on the target domain and the pre-trained model from the source domains. 
%final solution with the initial model.
They find that a simple $L^2$ regularization performs best.
SpotTune~\cite{guo2019spottune} proposes an adaptive fine-tuning approach  which aims to find the optimal fine-tuning strategy for each instance in the target domain. A policy network is learned
to make routing decisions on whether to pass through the fine-tuned layers or through the pre-trained layers.
More recently, PeterRec\cite{yuan2020parameter} keeps the pre-trained parameters unaltered, while introduces several network layers with learnable parameters.
%during fine-tuning by injecting a series of re-learned neural networks. 

Our work is inspired by \cite{guo2019spottune}.
However, we mainly focus on exploring the parameter transferability in different components of deep CTR models which have significantly different input features and structures compared with models in computer vision. Such differences bring challenges when we transfer deep CTR models from source domains to the target domain, as stated in Section~\ref{sec:intro}.

\subsection{Cross Domain Recommendation}
There are some work for cross domain recommendation that are relevant to our work. For example, CST \cite{DBLP:conf/aaai/PanXLY10}, NATR \cite{DBLP:conf/www/GaoCFZ00J19}, DARec\cite{DBLP:conf/ijcai/YuanYB19} focus on transferring feature representations, such as user/item embeddings, from source domains to target domains. Those methods are similar to model-based methods if we take the embeddings as part of the CTR model. However, these methods do not tell us whether other parts of a CTR model are transferable or not. We compare AutoFT with NATR in the experiments. Approaches such as CST and DARec that rely on the availability of rating matrix is beyond the scope of this work.

\section{Methodology}

\begin{figure*}[h]
\centering
\includegraphics[width=0.8\textwidth]{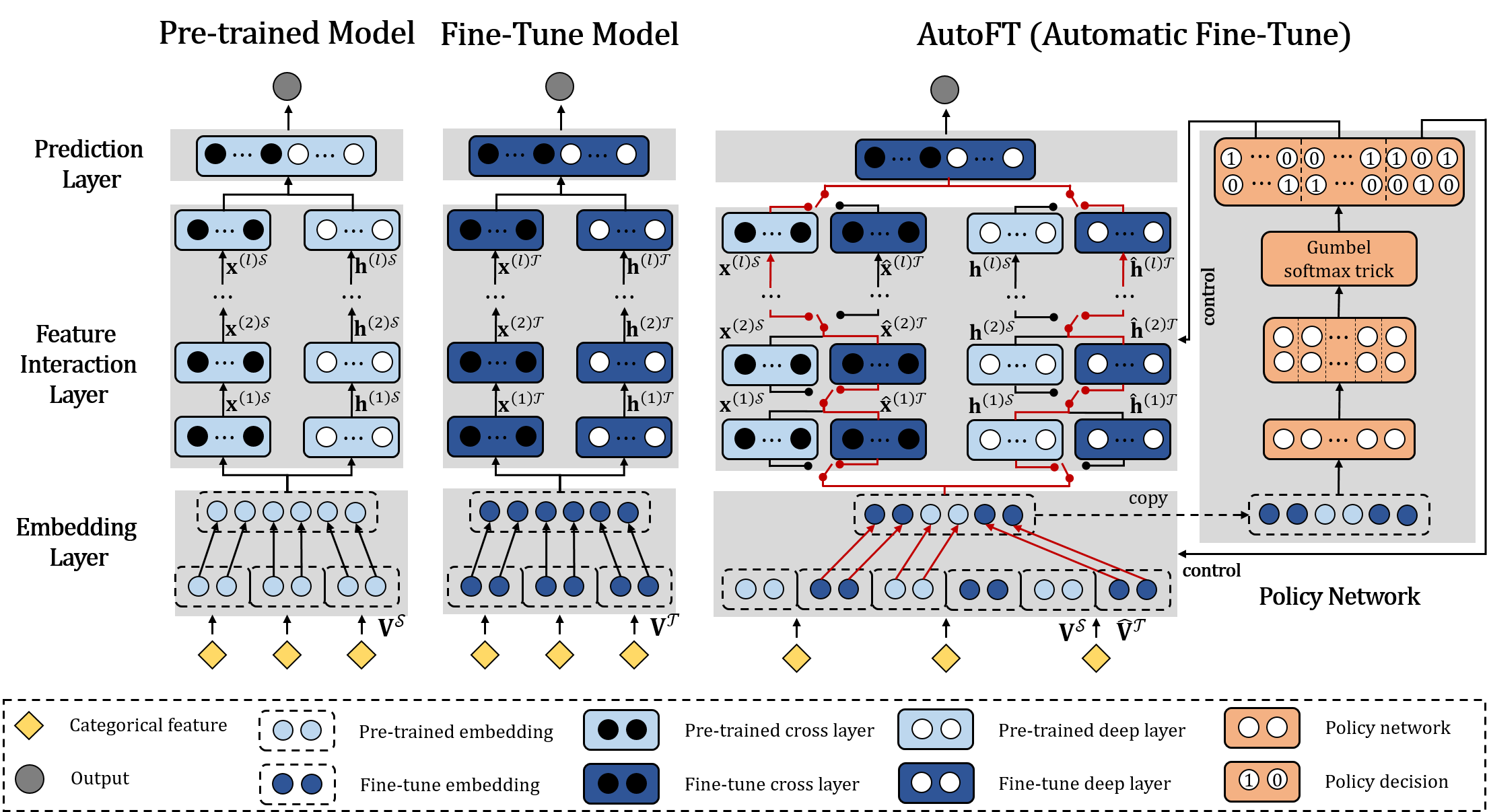} 
\caption{Framework of AutoFT compared to traditional pre-train and fine-tune strategies. Noted that the fine-tune parts share same color in Fine-Tune Model and AutoFT, but they have different parameters because of different training way.} 
\label{fig:framework} 
\end{figure*}

In this section, we introduce our proposed method AutoFT, which is an automatic fine-tuning strategy that decides which layers of a deep CTR model and which input fields should be frozen or fine-tuned for each target instance. AutoFT is compatible to any deep CTR models. With the loss of generality and for the ease of presentation, we select a representative CTR model, i.e., Deep \& Cross Network (DCN)~\cite{DBLP:conf/kdd/WangFFW17} as the backbone model for our fine-tuning method AutoFT. For clarity, we start by introducing the backbone DCN model in Section~\ref{sec:dcn}. 
%The architecture of DCN and useful notations are introduced in Section 3.1. 
Then, the details of the proposed AutoFT framework are presented in Section 3.2. Finally, in Section 3.3, we elaborate learning the policy network in AutoFT by using the Gumbel Softmax trick.
%which provides a solution for sampling from a discrete distribution.

\subsection{Backbone DCN Model}\label{sec:dcn}
%AutorFT could be applied to different deep neural network architectures, we choose the DCN framework for easy reference.
As stated earlier, most deep CTR models follow an embedding \& feature interaction paradigm, which consists of embedding layer, feature interaction layers and prediction layer.
DCN, introduced by Wang et al.~\cite{DBLP:conf/kdd/WangFFW17}, also an instantiation of this paradigm. We present the details of DCN model in this section.
% is in this framework which includes an embedding layer (feature representation), cross layers and deep layers (feature interaction), and a combination layer (prediction).

\noindent\textbf{Embedding layer.}
%Different from CV and NLP tasks, 
In CTR prediction, one-hot or multi-hot encoding is used to represent an input instance, as in Eq.(\ref{equ:hot}), which includes multiple categorical fields, such as item category. 
\begin{equation}\label{equ:hot}
  \mathbf{x}_{\text{hot}} = [\mathbf{x}_{\text{hot}, 1}, \mathbf{x}_{\text{hot}, 2}, \cdots, \mathbf{x}_{\text{hot}, m}],
\end{equation}
where $m$ is the number of fields and $\mathbf{x}_{\text{hot}, i}$ is the one-hot or multi-hot encoding vector of the $i$-th field in this instance. Each field consists of possibly many feature values, therefore 
the one-hot encoding vector $\mathbf{x}_{\text{hot}}$ is a high-dimensional sparse binary vector. To reduce the dimensionality, the embedding layer transforms the binary one-hot vectors into dense low-dimensional vectors, as in Eq.(\ref{equ:emb}).

\begin{equation}\label{equ:emb}
  \mathbf{x}_{i} = \mathbf{V}_{i} \mathbf{x}_{\text{hot}, i},
\end{equation}
% where $\mathbf{x}_{i}$ is the embedding vector of $\mathbf{x}_{\text{hot}, i}$, $\mathbf{V}_{i} \in \mathbb{R}^{{k \times n_{i}}}$ is the corresponding embedding matrix of the $i$-th field, and $k, n_{i}$ are the embedding size and number of features in the $i$-th field. The embedding matrices of all the fields are optimized together with network parameters.

where $\mathbf{x}_{i}$ is the embedding vector of $\mathbf{x}_{\text{hot}, i}$, $\mathbf{V}_{i} \in \mathbb{R}^{k \times n_{i}}$ is the corresponding embedding
matrix of the $i$-th field, and $k, n_{i}$ are the embedding size and number of features in the $i$-th field. The embedding matrices of all the fields are optimized together with network parameters.
%that normally will be optimized during training process.

%The representation of a one-hot vector $\mathbf{x}_{\text{hot}, i}$ with $j$-th element $\mathbf{x}_i[j] = 1$ is $\mathbf{x}_{i}$\textcolor{red}{???}. 
For a multivariate field, the representation of the multi-hot vector $\mathbf{x}_{\text{hot}, i}$ with $k$ elements $\mathbf{x}_i[j] = 1~(j = i_{1}, i_{2}, \cdots, i_{k})$ is the average of the individual embeddings:
\begin{equation}
  \mathbf{x}_{i} = \frac{1}{k}\sum_{j=i_{1}}^{i_{k}}\mathbf{V}_{i}\mathbf{x}_{\text{hot},i}[j].
\end{equation}
The final output of the embedding layer is the concatenation
of embedding vectors of all the fields:
\begin{equation}
  \mathbf{x} = [\mathbf{x}_{1}, \mathbf{x}_{2}, \cdots, \mathbf{x}_{m}].
\end{equation}

\noindent\textbf{Feature Interaction layers.}
The key challenge of CTR prediction is to model feature interactions effectively. 
In DCN model, the feature interaction layers consist of a \emph{cross network} and a \emph{deep network}. 

The cross network applies explicit feature interactions in an efficient way.
It consists of multiple layers, where each layer produces higher-order interactions based on existing ones, and keeps the interactions from previous layers. 
Let $\mathbf{x}^{(l)}$ denote the $l$-th cross layer. The output cross layer $\mathbf{x}^{(l+1)}$ is generated as in Eq.(\ref{equ:crosslayer}).
\begin{equation}\label{equ:crosslayer}
  \mathbf{x}^{(l+1)} = \mathbf{x}^{(0)} \mathbf{x}^{(l)\text{T}} \mathbf{w}_{\text{c}}^{(l)} + \mathbf{b}_{\text{c}}^{(l)} + \mathbf{x}^{(l)},
\end{equation}
where $\mathbf{w}_{\text{c}}^{(l)}$ and $\mathbf{b}_{\text{c}}^{(l)}$ are the weight and bias terms of the $l$-th cross layer, and $\mathbf{x}^{(0)}$ is first layer of cross network (which is equivalent to the output of embedding layer).

The deep network, in parallel with the cross network, is a fully-connected neural network, introduced to capture high-order nonlinear feature interactions. Let $\mathbf{h}^{(l)}$ denote the $l$-th deep layer, the output deep layer $\mathbf{h}^{(l+1)}$ is shown in Eq.(\ref{equ:deeplayer}).
\begin{equation}\label{equ:deeplayer}
  \mathbf{h}^{(l+1)} = \text{ReLU} (\mathbf{w}_{\text{d}}^{(l)} \mathbf{h}^{(l)} + \mathbf{b}_{\text{d}}^{(l)}), 
\end{equation}
where $\mathbf{w}_{\text{d}}^{(l)}$ and $\mathbf{b}_{\text{d}}^{(l)}$ are weight and bias terms for the $l$-th deep layer. ReLU is an activation function.

% Thus, the cross and deep networks learn both low- and high-order feature interactions.

\noindent\textbf{Prediction layer.}
The prediction layer combines the outputs from the cross network and the deep network, i.e., $\mathbf{x}^{(L_{\text{c}})}$ and $\mathbf{h}^{(L_{\text{d}})}$ to make prediction, where $L_{\text{c}}$ and $L_{\text{d}}$ represent the number of layers in the cross network and the deep network.
More specifically, the prediction is based on the concatenation of the two output vectors as in Eq.(\ref{equ:pred}).
\begin{equation}\label{equ:pred}
  \hat{y} = \text{Sigmoid}(\mathbf{w}_{\text{o}}[\mathbf{x}^{(L_{\text{c}})}, \mathbf{h}^{(L_{\text{d}})}] + \mathbf{b}_{\text{o}}),
\end{equation}
where $\mathbf{w}_{\text{o}}, \mathbf{b}_{\text{o}}$ are the weight and bias terms for the prediction layer. Sigmoid is an activation function.

The loss function is cross-entropy of the predicted values $\hat{y}$ and the labels $y$ with a regularization term,
\begin{equation}\label{equ:dcnloss}
  \mathcal{L}(y,\hat{y}) = -y\log(\hat{y}) - (1-y)\log(1-\hat{y}) + \lambda \lVert \mathbb{W} \rVert ^2,
\end{equation}
where $\mathbb{W}$ represents the set of parameters of DCN model, and $\lambda$ is a hyper-parameter to balance the prediction error and the regularization term.
% $\mathbb{W}$ represents the parameters of DCN model.

\subsection{AutoFT Framework}

The key idea of AutoFT is to train a set of learnable policy networks, one of which is to decide which fields of the input instance should be frozen or fine-tuned (referred as \textit{field-wise transfer policy}), while the others make routing decisions on whether to pass the input through the pre-trained network layers or the siamese fine-tuning network layers (referred as \textit{layer-wise transfer policy}). 
%Given a pre-trained model from source domains, the policy network will decide which layers of the pre-trained model should be fine-tuned and which layers should have their parameters shared with the source domains during training. 
Following the definition of transfer learning~\cite{cao2010adaptive}, the source domains are defined as $\mathcal{S}$ and the target domain is represented as $\mathcal{T}$.
%To tackle the transfer learning issue, the fine-tune technique have been widely used, which is fine-tuning all the layers based on target domains by using the pre-traiend parameters on source domains as a starting point. 
The key notations of the pre-trained and fine-tuned networks are shown in Tabel \ref{tbl:notations}. 

\begin{table}[t]
\centering
\caption{Related notations of the pre-trained and fine-tune model.}
\label{tbl:notations}
    \small
    \begin{tabular}{|l|l|} 
    \hline 
    Notation & \multicolumn{1}{c|}{Meaning} \\
    \hline  
    $\mathbf{V}^{\mathcal{S}}$ & pre-trained embedding matrix in source domain \\
    $\mathbf{V}^{\mathcal{T}}$ & fine-tuned embedding matrix in target domain \\
    $\mathbf{x}^{\mathcal{S}}$ & embedding vector of a source instance \\
    $\mathbf{x}^{\mathcal{T}}$ & embedding vector of a target instance \\
    \multirow{2}*{$\mathbf{w}_{\text{c}}^{\mathcal{S}}$, $\mathbf{b}_{\text{c}}^{\mathcal{S}}$} & pre-trained weight and bias terms of the cross \\ 
    & network in source domain \\
    \multirow{2}*{$\mathbf{w}_{\text{c}}^{\mathcal{T}}$, $\mathbf{b}_{\text{c}}^{\mathcal{T}}$} & fine-tuned weight and bias terms of the cross \\
    & network in target domain \\
    $\mathbf{x}^{(l)\mathcal{S}}$ & $l$-th cross layer of DCN model in source domain \\
    $\mathbf{x}^{(l)\mathcal{T}}$ & $l$-th cross layer of DCN model in target domain \\
    \multirow{2}*{$\mathbf{w}_{\text{d}}^{\mathcal{S}}$, $\mathbf{b}_{\text{d}}^{\mathcal{S}}$} & pre-trained weight and bias terms of the deep \\
    & network in source domain \\
    \multirow{2}*{$\mathbf{w}_{\text{d}}^{\mathcal{T}}$, $\mathbf{b}_{\text{d}}^{\mathcal{T}}$} & fine-tuned weight and bias terms of the deep \\
    & network in target domain \\
    $\mathbf{h}^{(l)\mathcal{S}}$ & $l$-th deep layer of DCN model in source domain \\
    $\mathbf{h}^{(l)\mathcal{T}}$ & $l$-th deep layer of DCN model in target domain \\
    \multirow{2}*{$\mathbb{W}^{\mathcal{S}}$} & parameters of pre-trained DCN neural network \\
    & in source domain\\
    \multirow{2}*{$\mathbb{W}^{\mathcal{T}}$} & parameters of fine-tuned DCN neural network \\
    & in target domain\\
    \hline 
    \end{tabular}
\end{table}

Figure \ref{fig:framework} presents the pre-trained CTR model, the fine-tuned CTR model and our proposed AutoFT framework. 
The pre-trained CTR model is trained based on the data from source domains or based on the data from all domains.
The fine-tuned CTR model is initialized by the pre-trained CTR model and fine-tuned with data from the target domain.
AutoFT involves three sets of parameters: the light-blue boxes contain parameters from the pre-trained CTR
model, the dark-blue boxes contain parameters initialized by the pre-trained model and will be fine-tuned during training and the red boxes contain the parameters of the policy networks. 
Compared to the original DCN network, AutoFT includes both the pre-trained and fine-tuned network, which has a similar structure as deep structured semantic models~\cite{huang2013learning}. 
There are three policy networks in AutoFT.
They decide how parameters are transferred in the embedding layer, the cross network and deep network, respectively (\textbf{for addressing challenge C2}).
Each policy network is a lightweight neural network, the input of which is the embedding of each instance such that each instance has its own fine-tune strategy (\textbf{for addressing challenge C3}). 
The output $\mathbf{p}$ of each policy network is a vector of binary values which serves as the indicator to choose the pre-trained parameters or the fine-tuned parameters.
\begin{equation}\label{equ:policy}
    \mathbf{p} = G(\text{ReLU}(\mathbf{W_p^2}\text{ReLU}(\mathbf{W_p^1}\mathbf{x}+\mathbf{b_p^1})+\mathbf{b_p^2}))
\end{equation}
Eq.(\ref{equ:policy}) shows an example of the policy network with two fully-connected layers and a Gumbel-Softmax trick $G$. 
Details of the Gumbel-Softmax trick will be illustrated in Section \ref{sec:gumbel}.
Other network structures are also possible for the policy networks.
% such as inner product and outer product are also adoptable.

\noindent\textbf{Field-wise transfer policy.} For the embedding layer, the policy network takes the pre-trained embedding $\mathbf{x}^{\mathcal{S}}$ from source domains as input and outputs a \emph{binary} vector $p_{\text{e}}(\mathbf{x}^{\mathcal{S}})$.
The $i$-th value in this vector, denoted as $P_{\text{e}}^{(i)}(\mathbf{x}^{\mathcal{S}})$ in Eq.(\ref{equ:embedpolicy}), decides whether the $i$-th field of features should use pre-trained embedding parameters $\mathbf{V}^{\mathcal{S}}$ or the fine-tuned embedding parameters $\hat{\mathbf{V}}^{\mathcal{T}}$. $\hat{\mathbf{x}}_{i}^{\mathcal{T}}$ represents feature embeddings where each field is selected either from pre-trained parameters or from fine-tuned parameters, controlled by the policy network.

\begin{equation}\label{equ:embedpolicy}  
  \hat{\mathbf{x}}_{i}^{\mathcal{T}} = P_{\text{e}}^{(i)}(\mathbf{x}^{\mathcal{S}}) \mathbf{V}_{i}^{\mathcal{S}} \mathbf{x}_{\text{hot}, i} + (1 - P_{\text{e}}^{(i)}(\mathbf{x}^{\mathcal{S}})) \hat{\mathbf{V}}_{i}^{\mathcal{T}} \mathbf{x}_{\text{hot}, i}.
\end{equation}
Noted that $\hat{\mathbf{V}}^{\mathcal{T}}$ and $\mathbf{V}^{\mathcal{T}}$ are different embedding matrices. 
$\mathbf{V}^{\mathcal{T}}$ is the embedding matrix in Fine-Tune model while $\hat{\mathbf{V}}^{\mathcal{T}}$ is the embedding matrix in AutoFT as shown in Figure~\ref{fig:framework}.

\textbf{Layer-wise transfer policy.} For the cross network and deep network, the input of the policy network is $\hat{\mathbf{x}}^{\mathcal{T}}$ computed based on Eq.(\ref{equ:embedpolicy}). 
The outputs are two binary vectors $P_{\text{c}}(\hat{\mathbf{x}}^{\mathcal{T}})$ and $P_{\text{d}}(\hat{\mathbf{x}}^{\mathcal{T}})$, to decide a ``route'' from the embedding layer to the prediction layer in deep network and cross network, respectively.
Each binary value in each of two vectors determines the direction of each step in the route, by specifying whether the parameters of the current layer should inherit from the pre-trained model or from the fine-tuned model, as shown in Figure \ref{fig:framework}. 

Specifically, in the cross network, the output of $l$-th cross layer is generated as in Eq.(\ref{equ:crosspolicy}).
\begin{equation}\label{equ:crosspolicy}
  \hat{\mathbf{x}}^{(l+1)\mathcal{T}} = P_{\text{c}}^{(l)}(\hat{\mathbf{x}}^{\mathcal{T}}) F_{\text{c}}(\hat{\mathbf{x}}^{(l)\mathcal{T}})
  + (1 - P_{\text{c}}^{(l)}(\hat{\mathbf{x}}^{\mathcal{T}}))
  \hat{F}_{\text{c}}(\hat{\mathbf{x}}^{(l)\mathcal{T}}).
\end{equation}
$F_{\text{c}}(\cdot)$ is a cross layer formulated in Eq.(\ref{equ:crosslayer}), which take $\mathbf{x}^{(l)}$ as input and outputs $\mathbf{x}^{(l+1)}$ with frozen pre-trained parameters $\mathbf{w}_{\text{c}}^{\mathcal{S}}$ and $\mathbf{b}_{\text{c}}^{\mathcal{S}}$. 
Similarly, $\hat{F}_{\text{c}}(\cdot)$ is a cross layer with the fine-tuned parameters $\hat{\mathbf{w}}_{\text{c}}^{\mathcal{T}}$ and $\hat{\mathbf{b}}_{\text{c}}^{\mathcal{T}}$. 

In the deep network, the output of the $l$-th deep layer $\hat{\mathbf{h}}^{(l+1)\mathcal{T}}$ is generated as in Eq.(\ref{equ:deeppolicy}).
\begin{equation}\label{equ:deeppolicy}
  \hat{\mathbf{h}}^{(l+1)\mathcal{T}} = P_{\text{d}}^{(l)}(\hat{\mathbf{x}}^{\mathcal{T}}) F_{\text{d}}(\hat{\mathbf{h}}^{(l)\mathcal{T}})
  + (1 - P_{\text{d}}^{(l)}(\hat{\mathbf{x}}^{\mathcal{T}}))
  \hat{F}_{\text{d}}(\hat{\mathbf{h}}^{(l)\mathcal{T}}).
\end{equation}
$F_{\text{d}}(\cdot)$ is a deep layer defined in Eq.(\ref{equ:deeplayer}) with frozen pre-trained parameters $\mathbf{w}_{\text{d}}^{\mathcal{S}}$ and $\mathbf{b}_{\text{d}}^{\mathcal{S}}$. 
Similarly, $\hat{F}_{\text{d}}(\cdot)$ is a deep layer with the fine-tuned parameters $\hat{\mathbf{w}}_{\text{d}}^{\mathcal{T}}$ and $\hat{\mathbf{b}}_{\text{d}}^{\mathcal{T}}$.

% \begin{figure}[h]
% \centering
% \includegraphics[width=0.48\textwidth]{images/fw2.png} 
% \caption{Introduction\textcolor{red}{change title and name}} 
% \label{Fig.Intro1} 
% \end{figure}

% As we shown in Figure \ref{Fig.Intro1}, the frozen part $F(\cdot)$ trained on the source domain is at the left side and the fine-tuned part $\hat{F}(\cdot)$ is on the right, which is optimized towards the target dataset from the replicated parameters of the frozen part. $P(\cdot)$ is the output of a lightweight policy network with two categories (freeze or finetune). (continuous or 01, reason of 01, difficulty).

The prediction layer combines the outputs of the cross network and the deep network to predict the click probability. 
It has been proven by~\cite{yosinski2014transferable, yuan2020parameter} that the prediction learns features that are specific to a particular task.
Thus, we always fine-tune the parameters of the prediction layer.

The loss function is the same as the one in Eq.(\ref{equ:dcnloss}).
During the training phase, the parameters from the pre-trained model are frozen, while other parameters are trainable.
During the inference phase, the policy networks generate a route from the embedding layer to the prediction layer, where each step of the route passes by either the pre-trained model or the fine-tuned model.
Since the policy networks are lightweight networks, the inference time of AutoFT is almost the same as DCN.

\subsection{Gumbel-Softmax Trick}\label{sec:gumbel}
AutoFT makes binary decisions at the embedding layer, the feature interaction layers (i.e., the cross network and deep network), to choose to inherit parameters from the pre-trained model or from the fine-tuned model. Such binary decisions are made by policy networks.
The output of the policy networks are discrete values which make the parameters in the policy networks non-differentiable. Therefore, it is difficult to optimize the network with back propagation. 
%There are several ways that allow us to “back-propagate” through the discrete nodes. 
The Gumbel-Softmax trick \cite{jang2016categoricalgumbel} is an approach to circumvent this problem. This trick provides a simple and effective way to draw samples from a categorical distribution parameterized by $\{\alpha_{0},\alpha_{2},\cdots,\alpha_{z-1}\}$, where $\alpha_{i}$ is the probability of category $i$, and $z$ is the number of categories. In this paper, there are two categories, $\alpha_{0}$ and $\alpha_{1}$, which are probabilities of using the frozen pre-trained parameters or the fine-tuned parameters.

During the forward pass, 
$z$ samples $G_0,\cdots,G_{z-1}$ are sampled from a standard Gumbel distribution $G = -log(-log(u))$, where $u$ is sampled from
a uniform distribution, i.e., $u \sim U[0,1]$.
%The Gumbel-Max trick is to draw samples from a discrete distribution parameterized by $\alpha_{i}$ in the following way.
Then each element in a policy vector $\mathbf{p}$ (corresponding to Eq.(\ref{equ:policy})) is computed by Eq.(\ref{equ:argmaxag}), where $\alpha_i$ is the output of the fully-connected layers in the policy networks.
\begin{equation}\label{equ:argmaxag}
  p_i=\mathop{\arg\max}_{i}[\log\alpha_{i}+G_{i}] \quad \text{for } i=0,\cdots,z-1
\end{equation}
%Suppose there are $l$ layers in the pre-trained model. The output of the policy network $\beta$ is a two-dimensional matrix, $\beta \in \mathbb{R}^{l\times2}$. Each row of $\beta$ represents the logits of a Gumbel-Max Distribution with two categories.

The argmax in Eq.(\ref{equ:argmaxag}) is non-differentiable.
Thus, during the backward pass, we replace the non-differentiable
sample from a categorical distribution with a differentiable sample
from a novel Gumbel-Softmax distribution which can be smoothly
annealed into a categorical distribution.
The gradient of the discrete samples are approximated by computing the gradient of the continuous softmax relaxation as in Eq.(\ref{equ:softrelax}).

%We can use the \textit{Gumbel-Softmax distribution}, which adopts softmax as a continuous relaxation to argmax.

%We represent $X$ as a one-hot vector where the index of the non-zero entry of the vector is equal to $X$, and relax the one-hot encoding of $X$ to a $z$-dimensional real-valued vector $Y$ using softmax:

\begin{equation}\label{equ:softrelax}
  Y_{i}=\frac{\exp((\log \alpha_{i} + G_{i})/\tau)}{\sum_{j=1}^{z}\exp((\log \alpha_{j} + G_{j})/\tau)} \quad \text{for } i=0,\cdots,z-1
\end{equation}
where $\tau$ is a temperature parameter, which controls the discreteness of the output vector $Y$. When $\tau$ becomes closer to 0, the samples from the Gumbel-Softmax distribution become indistinguishable from the discrete distribution.  % (i.e, almost the same as the one-hot vector).
The Gumbel-Softmax distribution is smooth for $\tau > 0$ and therefore has well-defined gradients with respect to the parameters $\alpha_{i}$.

By using a standard classification loss for the target domain, the policy network is jointly trained with the fine-tuned parameters to find the optimal fine-tune strategy that maximizes the accuracy of CTR prediction in the target domain.
This whole process is illustrated in Figure \ref{fig:framework}.
%Similar to other work, we generate all freeze/fine-tune decisions for all residual blocks at once, instead of relying on features of intermediate layers of the pre-trained model to obtain the fine-tuning policy. More specifically, 

\section{Experiments}

In order to verify the effectiveness of the proposed framework AutoFT, we conduct extensive experiments on two public benchmark datasets and one industrial dataset.
We mainly focus on the following research questions (RQs).
\begin{itemize}
    \item RQ1: Is AutoFT able to positively transfer the pre-trained model to the target domain, can it outperform state-of-the-art approaches?
    \item RQ2: How do different components of proposed AutoFT, i.e., field-wise transfer policy and layer-wise transfer policy, contribute to the performance?
    \item RQ3: Is the policy based learning framework AutoFT compatible with any deep CTR models, such as DNN and DeepFM?
    \item RQ4: Which layers of the deep CTR models need more fine-tuning?
\end{itemize}

\subsection{Datasets}
Experiments are conducted on the following two public datasets (MovieLens-1M and Amazon review data) and one private industrial dataset.
The statistics of the datasets are in Table \ref{tbl:statdataset}.

\begin{table}[h]
\centering
\small
\caption{Statistics of the datasets. We show the number of instances in training, validation, and testing data from source and target domains, respectively, and other important details.}
%\textcolor{blue}{the first table adds the datasets column?}
\label{tbl:statdataset}
    % \begin{tabular}{|l|l|c|c|c|c} 
    % \hline 
    % &Domains & Training & Validation & Test \\
    % \hline  
    % Source & Male users  & 602975 & 75263 & 75531\\
    % Target & Female users & 197025 & 24737 & 24678\\
    % \hline 
    % Source & \textit{Toys and Games} & 6560984 & 8201231 & 8201231\\
    % Target & \textit{Video Games}    & 2052279 & 513069 & 513069\\
    % \hline
    % Source & \textit{Must-have Apps} & 68129281 & 9264471 & 9259634\\
    % Target & \textit{Novel-Fun}  & 4296836 & 622236 & 632941\\
    % \hline
    % \end{tabular}
    
    \begin{tabular}{|l|c|c|c|} 
    \hline 
    Datasets & MovieLens & Amazon & Industrial data \\
    \hline
    Source training    & 602,975 & 6,560,984 & 68,129,281 \\ 
    Target training   & 197,025 & 2,052,279 & 4,296,836 \\
    \hline
    Source validation  & 75,263  & 8,201,231 & 9,264,471 \\
    Target validation  & 24,737  & 513,069  & 622,236 \\
    \hline
    Source testing    & 75,531  & 8,201,231 & 9,259,634 \\
    Target testing    & 24,678  & 513,069  & 632,941 \\\hline
    Ratio of positive   & 0.5753 & 0.6040 & 0.0223 \\
    \hline  
    Users' overlap rate & 0.0\%   & 0.3\% & 1.7\% \\
    Items' overlap rate & 100.0\% & 0.0\% & 78.6\% \\
    \hline
    \end{tabular}
\end{table}

% \begin{table}[h]
% \centering
% \small
% \caption{Statistics of the used datasets.}
% \label{tbl:statdataset}
%     \begin{tabular}{|l|c|c|c|} 
%     \hline 
%     Datasets & MovieLens & Amazon & Industrial data \\
%     \hline  
%     Users' overlap rate & 0.0\%   & 0.3\% &  \\
%     Items' overlap rate & 100.0\% & 0.0\% &  \\
%     Ratio of positive   & 0.5753 &  &  \\
%     \hline
%     \end{tabular}
% \end{table}

\textbf{MovieLens-1M dataset}\footnote{https://www.grouplens.org/datasets/movielens/}: The dataset includes 1 million movie rating records over thousands of movies and users \cite{harper2015movielens}. The features we used are user id, gender, occupation, user behavior, movie id, release year, and movie genres. 
%We first binarize ratings to simulate the CTR prediction task, i.e., 
The ratings greater than 3 are taken as positive samples, and others are taken as negative samples \cite{volkovs2015effective}. 
The dataset is randomly divided into the training, validation, and test sets with a ratio of 8:1:1. As we analysed in the data, the female and male users have different preferences and click behaviours about movies. Thus, we regard male as the source domain and female as the target domain.
 
\textbf{Amazon review data}\footnote{https://www.nijianmo.github.io/amazon/index.html}: This dataset contains product reviews and metadata from Amazon, including 24 categories spanning from 1996 to 2018 \cite{ni2019justifying}. The data we used are $5$-cores records of products in \textit{Toys and Games} and \textit{Video Games}.
%These data have been reduced from complete data, such that 
Each of the users and items have 5 reviews, respectively. Different from MoviLens-1M dataset, we take ratings greater than 4 as positive samples to balance the positive samples and negative samples. The features we used are reviewerID, asin, vote, style and reviewTime. The training, validation, and test sets are also divided randomly with a ratio of 8:1:1. 
%Normally, the categories with rich records are chosen as source domain to learn more general transferred knowledge. 
Records of \textit{Toys and Games} are used as source domain since this category has more records than \textit{Video Games}, which is more reasonable to transfer general knowledge.

\textbf{Industrial dataset}: This dataset is collected from a real commercial recommender system deployed in the company X's App Store. The dataset contains app features (e.g., ID, category), user features (e.g., user’s click behavior history) and context features (e.g., time).
We choose the two recommendation scenarios, \textit{Must-have Apps} and \textit{Novel-Fun} as the source domain and target domain, respectively. 
Each domain contains tens of thousands of apps and more than ten million users. We collect 73M records from the download logs, which represent the operations of users to apps. In training process, the source domain records is used to training the model with the target domain data be the validation and test dataset to evaluate the model. 

\subsection{Baselines \& experimental settings}
We compare AutoFT with multiple baseline methods to verify its effectiveness for CTR prediction.
All these methods are evaluated on the testing data from the target domain.

We firstly introduce three straightforward baseline methods that are easy to think of without transferring any parameters from the pre-trained model.  \textbf{Target-only}, Source, and All, which use \textbf{Source-only}, and \textbf{All} domains for training, respectively. 

\begin{itemize}
    \item \textbf{Target-only} is the method to use target domain data for both training and testing.
    \item \textbf{Source-only} is to use source domain data for training and use target domain data for testing.
    \item \textbf{All} is to use both source and target domain data for training, and use target domain data for testing.
\end{itemize}

Then we describe four representative state-of-the-art baseline approaches, i.e., Fine-tune, ProgressiveNN, $L^{2}$-SP and SpotTune, of transferring parameters from the pre-trained model.
%the following state-of-the-art fine-tuning and regularization techniques are used for comparison. For transfer parameters, we choose both the 
The pre-trained model is learned either from \textit{Source-only} or \textit{All} method.

\begin{itemize}
    \item \textbf{Fine-Tune} ~\cite{yosinski2014transferable} is the method to learn a model which is initialized by a pre-trained model. Then all the parameters of the model are fine-tuned using data from the target domain.
    \item \textbf{PTU} ~\cite{PTU} is a parameter transfer unit, that produces a weighted sum of the activations from both the already trained source domain network and the target domain network.
    \item \textbf{ProgressiveNN} ~\cite{rusu2016progressiveNN} proposes to aggregate the pre-trained parameters to the fine-tuning parameters with an element-wise non-linear function.
    %When switching from a source task to a target task, the pre-trained parameters are frozen and a new column with fine-tuning parameters is instantiated, where 
    The next layer receives input from both previous pre-trained and fine-tuning layers via lateral connections.
    \item \textbf{$L^{2}$-SP} \cite{xuhong2018explicit/L2SP} is a regularization-based method for fine-tuning. A simple $L^{2}$-penalty between the shared pre-trained network parameters $\mathbb{W}^{\mathcal{S}}$ and the fine-tuning parameters $\mathbb{W}^{\mathcal{T}}$ is applied to constrain the effective search space around the initial pre-trained model.
    \item \textbf{SpotTune} \cite{guo2019spottune} aims to dynamically route blocks in Residual Network model, which improves the classification performance in CV. For the CTR prediction tasks, this method is corresponding to our layer-wise policy network which is only used to control the choice in the cross and deep network.
    \item \textbf{NATR} ~\cite{DBLP:conf/www/GaoCFZ00J19} resolves the key challenges in leveraging transferred item embeddings. The two level attention design allows NATR to distill useful signal from transferred item embeddings, and appropriately combine them with the data of the target domain.
\end{itemize}

All methods are implemented with TensorFlow\footnote{https://www.tensorflow.org/} and trained on GPUs (Tesla P100) with Adam optimizer\cite{DBLP:journals/corr/KingmaB14}. 
All experiments are repeated five times and the average values are reported. For the hyper-parameters, we tune learning rate to be from 0.001 to 0.0001, set batch size to be 8000 and embedding dimension to be 80. Note that, DCN are used for backbone network in AutoFT. For network architecture, the cross network has 3 layers iteration, the deep network has 4 layers with dimensions specifically of 1024, 512, 256, 128. Finally, the common evaluation metrics, AUC (Area Under ROC) and Logloss (cross-entropy), are used for comparison.

\begin{table*}[ht]
\centering

\caption{Performance comparison of different methods. $^{*}$ denotes statistically significant improvement (measured by t-test with $p$-value$<$0.005) over baselines.}

\label{tbl:overallperformance}

\begin{tabular}{|l|c c|c c| c c|} 
\hline
\multirow{2}*{Method} & \multicolumn{2}{c|}{MovieLens-1M} & \multicolumn{2}{c|}{Amazon review data} & \multicolumn{2}{c|}{Industrial dataset} \\
\cline{2-7}
 & AUC & LogLoss & AUC & LogLoss & AUC & LogLoss \\
\hline  
Target-only           & 0.7561 & 0.6791 & 0.7120 & 0.6461 & 0.8949 & 0.03948 \\
\hline
\hline
Source-only & 0.7112  & 0.6701 & 0.6010 & 0.6549 & 0.8084 & 0.04723 \\
Source-Fine-Tune      & 0.7807 & 0.5609 & 0.7128 & 0.6007 & 0.9010 & 0.03907 \\
Source-ProgressiveNN  & \underline{0.7817} & \underline{0.5445} & 0.7124 & 0.6023 & \underline{0.9019} & \underline{0.03884} \\
Source-PTU  & 0.7809 & 0.5552 & 0.7130 & 0.5999 & 0.9013 & 0.03904 \\
Source-$L^2$-SP       & 0.7813 & 0.5508 & 0.7131 & 0.5999 & 0.9011 & 0.03905 \\
Source-SpotTune       & 0.7814 & 0.5506 & \underline{0.7133} & \underline{0.5997} & 0.9016 & 0.03885 \\
Source-NATR & ~~~~/ & ~~~~/ & ~~~~/ & ~~~~/  & 0.9014 & 0.03904 \\
\hline
Source-AutoFT   & \textbf{0.7822}$^{*}$ & \textbf{0.5504}$^{*}$ & \textbf{0.7138}$^{*}$ & \textbf{0.5968}$^{*}$ & \textbf{0.9024}$^{*}$ & \textbf{0.03882}$^{*}$ \\
\hline
\hline
All                   & 0.7776 & 0.6137 & 0.7106 & 0.6042 & 0.9039 & 0.03857 \\
All-Fine-Tune         & 0.7843 & 0.5430 & 0.7141 & 0.6004 & 0.9040 & 0.03855 \\
All-ProgressiveNN     & 0.7832 & 0.5437 & \underline{0.7146} & \underline{0.5965} & 0.9040 & 0.03852 \\
All-PTU  & 0.7863 & 0.5415 & 0.7152 & 0.5976 & 0.9039 & 0.03858 \\
All-$L^2$-SP          & 0.7864 & 0.5407 & 0.7143 & 0.6002 & 0.9046 & 0.03847 \\
All-SpotTune          & \underline{0.7885} & \underline{0.5398} & \underline{0.7146} & 0.5980 & 0.9044 & 0.03878 \\
All-NATR &  ~~~~/ &  ~~~~/ &  ~~~~/ &  ~~~~/ & \underline{0.9048} & \underline{0.03846} \\
\hline 
All-AutoFT      & \textbf{0.7932}$^{*}$ & \textbf{0.5375}$^{*}$ & \textbf{0.7166}$^{*}$ & \textbf{0.5945}$^{*}$ & \textbf{0.9052}$^{*}$ & \textbf{0.03843}$^{*}$ \\
\hline
\end{tabular}
\end{table*}

\subsection{Overall Performance (RQ1)}

The performance of the baseline approaches and our proposed framework AutoFT is shown in Table \ref{tbl:overallperformance}.
We divide the results into three groups.
The first group shows the results of the basic baseline method Target-only.
The second group of methods are based on a pre-trained model, i.e., Source-only, that is trained on data from source domains.
The third group of methods are based on a pre-trained model, i.e., All, that is trained on data from all domains.
The best results of the baseline methods in the second and third groups are underlined.
From this table, we have the following observations.

\textit{Firstly}, AutoFT performs best with respect to AUC among all the approaches on all datasets no matter that the pre-trained model is based on Source-only or based on All, which demonstrates the effectiveness of our proposed framework.
Moreover, the state-of-the-art baseline approaches such as ProgressiveNN, PYU, $L^2$-SP, and SpotTune, though do not dominate each other, generally performs better than the basic baseline approaches such as Target-only, Source-only, and All.
%, which is consist with the report in the literature [XXX].
However, their performance is inferior to AutoFT since they do not distinguish different component of a CTR model, thus, are less flexible when transferring different amount of valuable information for different components or instances.

\textit{Secondly}, we can observe that the approaches that include both source domains and target domain data perform better than Target-only and perform much better than Source-only.
This observation implies that importing extra data to the target domain can positively transfer information to improve the prediction accuracy.
For the scenario that share more common items, such as MovieLens-1M and Industrial data, the benefit of transferring is higher than that share less common items, such as Amazon Review.

\textit{Thirdly}, we found that approaches using All as the pre-trained model generally performs better than approaches using Source-only as the pre-trained model.
This is probably for the reason that All include more ``general'' features and parameters than Source-only, which indicates that a more ``general'' model that includes information from multiple domains that are relevant to the target domain can boost the performance of the target domain to a larger margin.

\subsection{Ablation study (RQ2)}
In this section, we study how field-wise transfer policy and layer-wise transfer policy, which are corresponding to different policy networks in AutoFT, contribute to the performance.
In the experiments, we only keep the policy network for the embedding layer and remove other policy networks from All-AutoFT. This method is denoted as \textit{Embedding} in Table \ref{tbl:ablation}.
Similarly, \textit{Cross}, \textit{Deep}, \textit{Cross \& Deep} represent the methods that only keep the policy networks for the cross layers, the deep layers and both the cross layers and deep layers, respectively.

From Table \ref{tbl:ablation}, we can observe that the performance of all variants drops when we remove certain policy networks from AutoFT which verifies that both field-wise transfer policy and layer-wise transfer policy play important roles in the performance.
For MovieLens-1M data, the drop of the performance with respect to AUC of \textit{Cross} and \textit{Deep} is significant compared with AutoFT when we remove field-wise policy network and one of the layer-wise policy networks.
For Amazon data, the drop of their performance is less significant. 
That implies for the dataset that shares more items in the source and target domains, the pre-trained model contains more general information that help the prediction in the target domain, which should be included by policy networks.
%For MovieLens-1M data, the drop of the performance with respect to AUC of \textit{Cross} and \textit{Deep} is significant compared with AutoFT when we remove field-wise policy network and one of the layer-wise policy networks.\textcolor{blue}{(describe more clearly)}
%For Amazon data, the drop of their performance is less significant. 
%That implies for the dataset that shares more items in the source and target domains, the pre-trained model contains more general information that help the prediction in the target domain, which should be included by policy networks.

\begin{table}[t]
\small
\caption{The performance of policy net of each component in DCN.}
\label{tbl:ablation}
\begin{tabular}{|l|c c|c c|} 
\hline
Policy-based & \multicolumn{2}{c|}{MovieLens-1M} & \multicolumn{2}{c|}{Amazon review}  \\
\cline{2-5}
Component & AUC & LogLoss & AUC & LogLoss \\
\hline 
All-Fine-Tune     &0.7843  &0.5430  & 0.7141 &  0.6004\\
\hline
Embedding     & 0.7878 & 0.5404 & 0.7135 & 0.5995 \\
Cross         & 0.7866 & 0.5412 & 0.7136 & 0.5990 \\
Deep          & 0.7854 & 0.5432 & \underline{0.7152} & \underline{0.5968} \\
Cross \& Deep & \underline{0.7885} & \underline{0.5398} & 0.7146 & 0.5980 \\
\hline
All-AutoFT        & \textbf{0.7932}  & \textbf{0.5375} & \textbf{0.7166} & \textbf{0.594}5 \\
\hline
\end{tabular}
\end{table}

\subsection{Compatibility of AutoFT (RQ3)}
In this section, we conduct experiments to verify whether AutoFT is compatible with different deep CTR models. We choose three representative CTR prediction models for our experiment, they are DCN, DNN and DeepFM. 
We keep the network architectures of deep layers of DCN and DNN to be the same which are of size 1024, 512, 256, 128.
We set the deep layers of DeepFM to be of size 400, 400, 400 to check if our proposed framework works for a different network size.
%with the policy network dimensions specifically of 512, 256, 128.
The pre-trained model is the baseline All. 
The experimental results are shown in Table \ref{tbl:compatibility}.
The difference between All-Fine-Tune and All-AutoFT is that All-Fine-Tune does not include any transfer policies, all its parameters are fine-tuned.
From Table \ref{tbl:compatibility}, we can observe that for various deep CTR models, AutoFT always achieves better results than the Fine-Tune method, which demonstrates the compatibility of our proposed framework.

\begin{table}[h]
\centering
\footnotesize
\caption{Compatibility of AutoFT with classic deep CTR models.}
\label{tbl:compatibility}
\begin{tabular}{|l|c c|c c|} 
\hline
\multirow{2}*{Model} & \multicolumn{2}{c|}{MovieLens-1M} & \multicolumn{2}{c|}{Amazon review}  \\
\cline{2-5}
& AUC & LogLoss & AUC & LogLoss \\
\hline  
\footnotesize{DCN Fine-Tune}   & 0.7843 & 0.5430 & 0.7141 & 0.6004 \\
\footnotesize{DCN AutoFT}      & \textbf{0.7932} & \textbf{0.5375} & \textbf{0.7166} & \textbf{0.5945} \\
\hline
\footnotesize{DNN Fine-Tune}   & 0.7754 & 0.6198 & 0.7131 & 0.6002 \\
\footnotesize{DNN AutoFT}      & \textbf{0.7840} & \textbf{0.5453} & \textbf{0.7145} & \textbf{0.5987} \\
\hline
\footnotesize{DeepFM Fine-Tune} & 0.7780 & 0.5844 & 0.7120 & 0.5997 \\
\footnotesize{DeepFM AutoFT}    & \textbf{0.7852} & \textbf{0.5434} & \textbf{0.7128} & \textbf{0.5991} \\
\hline
\end{tabular}
\end{table}

\subsection{Visualization of Policy (RQ4)}
\begin{figure}[h]
    \begin{minipage}[t]{0.49\linewidth}
    \centering
    \caption*{\small{MovieLens-1M}}\vspace{-12pt}
	\includegraphics[width=1.6in]{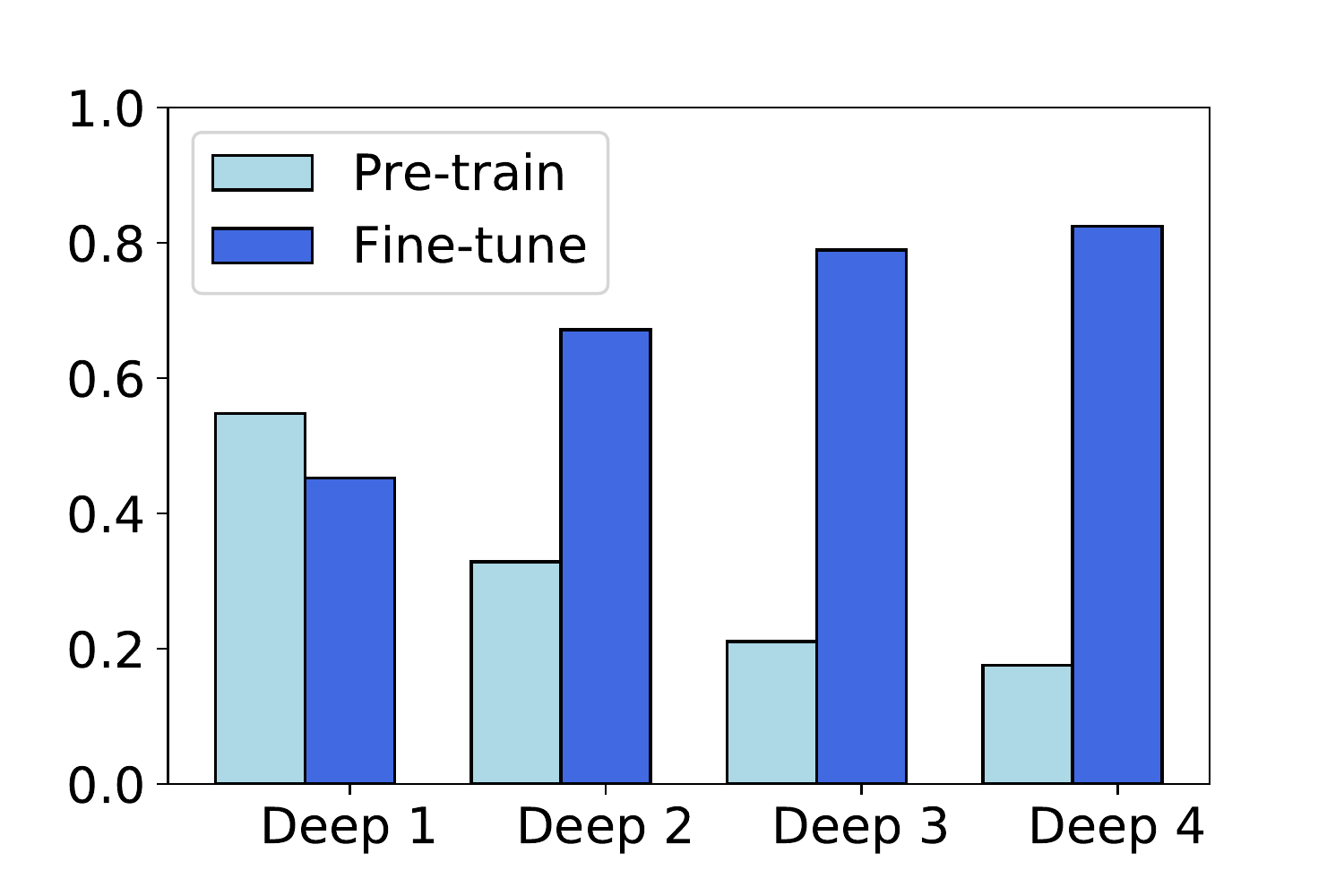}
	\includegraphics[width=1.6in]{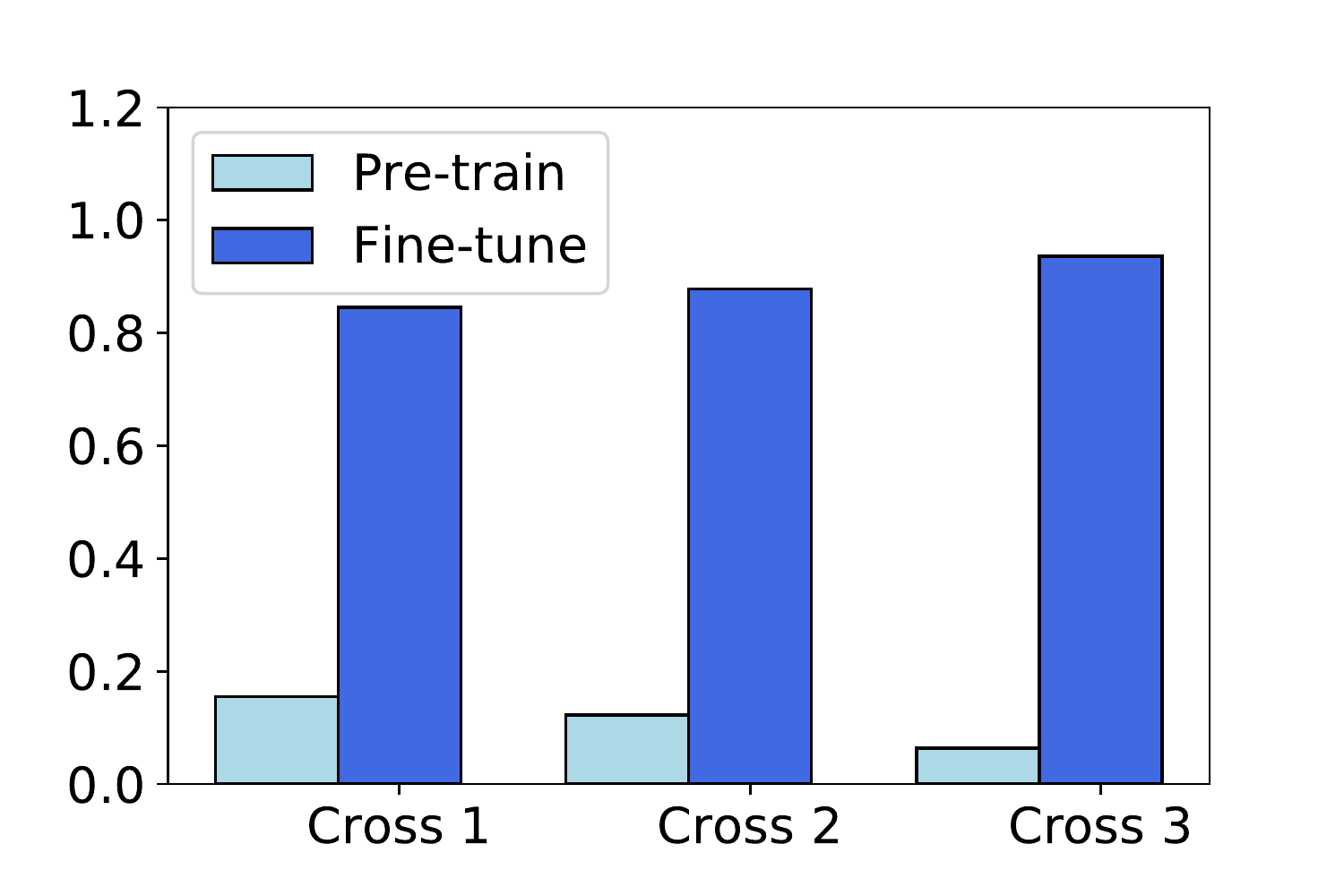}
    \end{minipage}
	%\hspace{0.2in}
    \begin{minipage}[t]{0.49\linewidth}
    \centering
    \caption*{\small{Amazon review}}\vspace{-12pt}
	\includegraphics[width=1.6in]{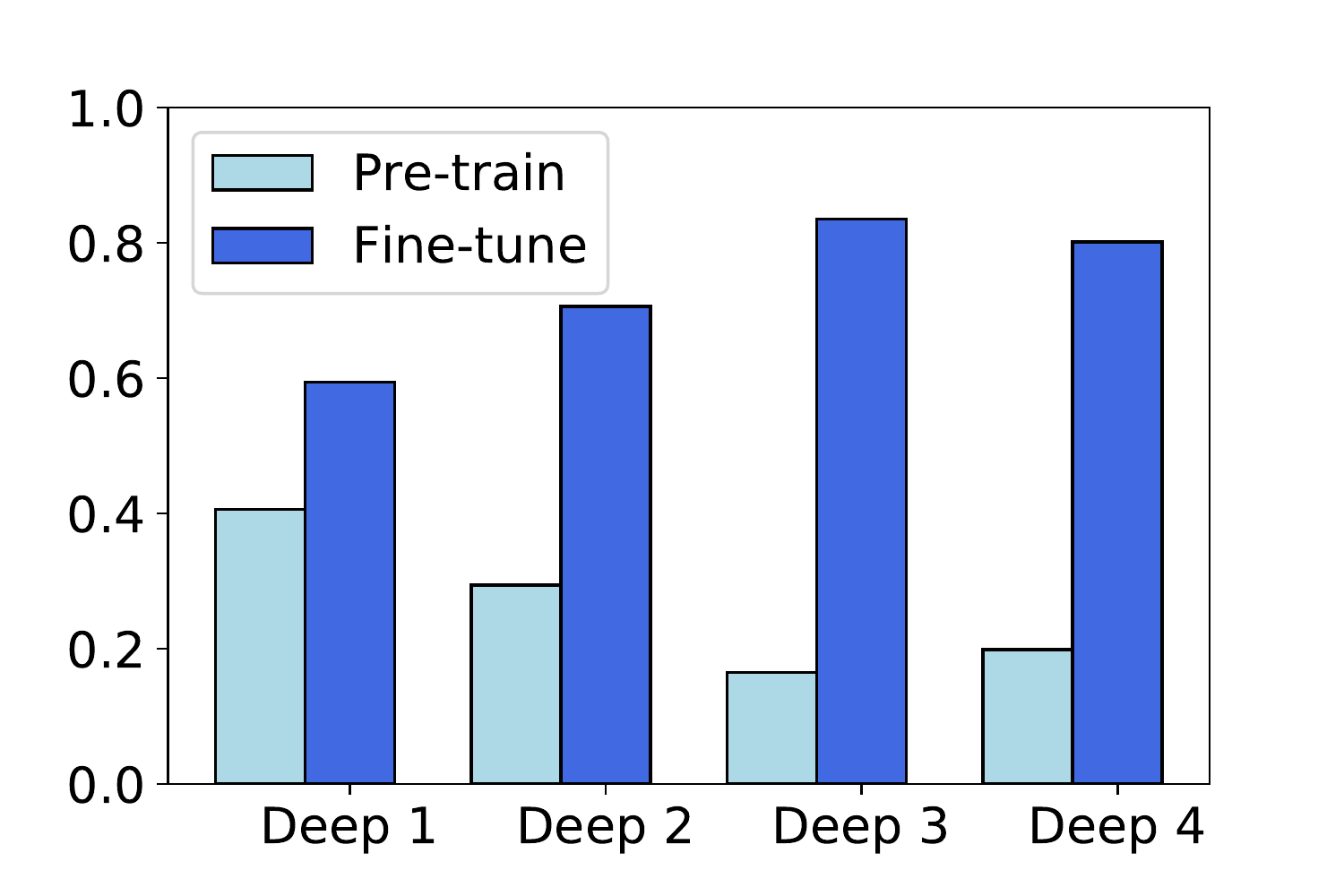}
	\includegraphics[width=1.6in]{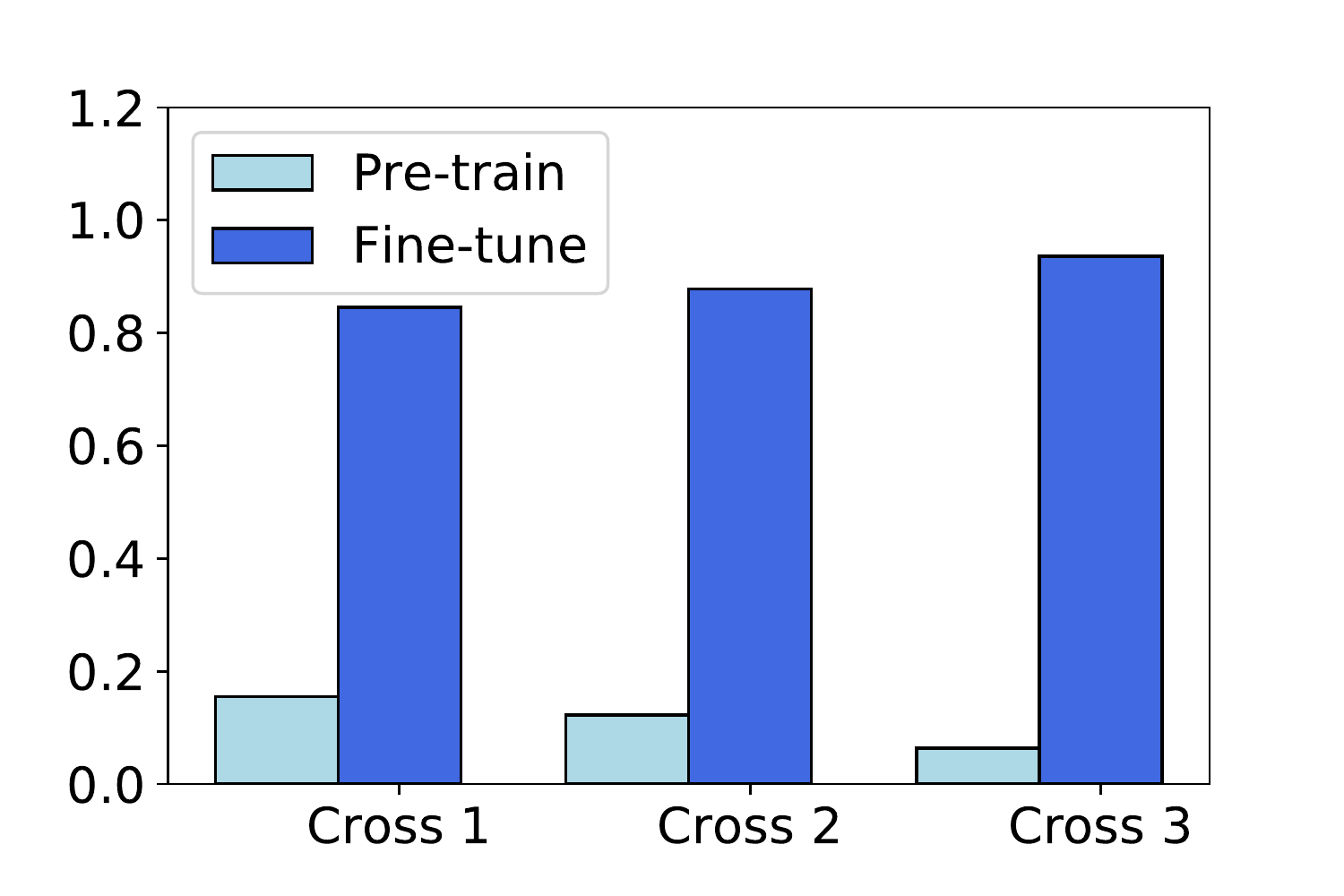}
    \end{minipage}
\caption{Visualization of AutoFT. The first row is about deep layers, and the second row is about the cross layers.} 
\label{fig:result} 
\end{figure}
To better understand the fine-tuning decisions learned by the policy networks, we visualize them on the two public benchmark datasets.
% The polices are learned on field-wise and layer-wise, however no interesting observations were found on field-wise visualisation.
For field-wise transfer policy, we count the number of instances that either using the pre-trained embedding or the fine-tuned embedding.
However, we do not find any special trend among different fields.
For layer-wise transfer policy, we counted each instance's layer-wise decision of using pre-trained or fine-tune parameters during testing. The percentage of the pre-trained and fine-tune policy in each layer based on MovieLens-1M and Amazon data are shown in Figure \ref{fig:result}. For example, a light-blue pre-train bar with a number of 0.2 means 20\% of the instances in the test target domain share the pre-trained parameters in this layer and the 80\% of instances use the fine-tuned parameters. 
The visualization shows that different datasets and different layers have specific fine-tuning policies, which validates our hypothesis that it is more accurate to have an instance-based fine-tuning policy for CTR prediction. AutoFT allows us to automatically learn the right policy for each dataset, as well as for each training instance, without much manual effort. By observing the percentages of pre-train and fine-tune bars in each sub-figures, we find that more and more instances choose the fine-tune direction on higher layers. This also verifies that the lower layers could learn more general features and the parameters in higher layers are more specific to each domains, which need to be fine-tuned during transfer learning.

\section{Conclusion}
In this paper, we propose a framework AutoFT to automatically incorporate parameters from a pre-trained model to facilitate the CTR prediction accuracy in a new target domain.
The field-wise policy in AutoFT decides which fields of embedding features should be inherited from the pre-trained model or should be fine-tuned.
The layer-wise policy generates route decisions between pre-trained and fine-tuned parameters for each layer in the deep and cross networks. 
To achieve this, lightweight policy networks are jointly trained with the target domain using a Gumbel-Softmax trick. 
The proposed framework is generally applicable to many deep CTR models. Extensive offline experiments demonstrate the superior performance of AutoFT.

% \clearpage
% \newpage
\bibliographystyle{named}
\bibliography{ijcai21}
\clearpage
\newpage

\end{document}